\documentclass[12pt]{iopart}
\usepackage{iopams}
\begin{document}
\title[Symmetry groups of single-wall carbon nanotubes]{On the linear representations of the symmetry   
              groups of single-wall carbon nanotubes}
\author{Nicolae Cotfas}
\address{Faculty of Physics, University of Bucharest,
PO Box 76-54, Post Office 76, Bucharest, Romania,\\ 
E-mail address: ncotfas@yahoo.com } 
\begin{abstract}
The positions of atoms forming a carbon nanotube are usually 
described by using a system of generators of the symmetry group. Each atomic position 
corresponds to an element of the set 
$ \mathbb{Z}\times \{ 0,1,...,n\} \times \{ 0,1\}$,
where $n$ depends on the considered nanotube.
We obtain an alternate rather different description by starting from 
a three-axes description of the honeycomb lattice.
In our mathematical model, which is a factor space defined by an 
equivalence relation in the set 
$ \left\{ (v_0,v_1,v_2)\in \mathbb{Z}^3\ |\ v_0+v_1+v_2\in \{ 0,1\}\right\}$,
the neighbours of an atomic position can be described in a simpler way,
and the mathematical objects with geometric or physical significance have
a simpler and more symmetric form. We present some results concerning the 
linear representations of single-wall carbon nanotubes in order to illustrate 
the proposed approach.
\end{abstract}
\maketitle

\section{Introduction}

A single-wall carbon nanotube is a cylindrical structure with a diameter of a 
few nanometers, periodic along its axis, which can be imagined as a rolled up 
honeycomb lattice. The high symmetry of carbon nanotubes has facilitated the 
theoretical investigation of the physical phenomena occurring in these materials \cite{A,Dre,H,J,S,W}.
The spatial symmetries (translations, rotations and screw axes, mirror and
glide planes, etc.) form a line group, which is the maximal subgroup of the
Euclidean group that leaves the nanotube invariant. The role of this group 
is analogous  to that of crystallographic space groups in solid state physics.
Some important properties of the band structure (electronic, phonon, etc.)
can be directly deduced from the symmetry groups.

The symmetry group of a carbon nanotube depends on the diameter of the tubule
and on the helical arrangement of the carbon hexagons.
The irreducible representations of these groups are well-known \cite{D3,D2,Vu}, but, generally, 
equivalent representations offer distinct formal advantages. A calculation very simple
in a representation can become much more complicated in an equivalent representation.
Therefore, we think it is worth looking for new representations, 
and for new ways to describe the atomic structure of these materials. 
Our aim is to present an improved version of the mathematical model proposed in \cite{C0}
and some applications illustrating this new approach.

\section{Honeycomb lattice in a three-axes description}

\noindent The vectors corresponding to the vertices of a regular triangle
\begin{equation}
 e_0=(1,0),\qquad e_2=\left( -\frac{1}{2},\frac{\sqrt{3}}{2}\right), \qquad
e_3=\left( -\frac{1}{2},-\frac{\sqrt{3}}{2}\right)  
\end{equation}
allow us to define the bijecton
\begin{equation}
 \mathcal{L}\longrightarrow \mathbb{L}:\ (v_0,v_1,v_2)\mapsto
v_0e_0+v_1e_1+v_2e_2 
\end{equation}
from the set
\begin{equation}
\mathcal{L}=\{ \ v=(v_0,v_1,v_2)\in \mathbb{Z}^3\ \ 
|\ \ v_0+v_1+v_2\in \{ 0;1\}\ \} 
\end{equation}
to the set \ $\mathbb{L}$ \ 
of all the vertices of a {\it honeycomb lattice.} 
The subset $\mathcal {L}$ of $\mathbb{Z}^3$ becomes in this way a 
mathematical model for the honeycomb lattice.
One can remark that
\begin{equation}
 \mathcal{L}=\mathcal{T}\cup (\mathcal{T}+(1,0,0))
\end{equation}
where
\begin{equation}
\mathcal{T}=\{ \ v=(v_0,v_1,v_2)\in \mathbb{Z}^3\ |\ v_0+v_1+v_2=0 \ \}. 
\end{equation}

The mapping 
\begin{equation}
 d:\mathcal{L}\times \mathcal{L}\longrightarrow \mathbb{N},\qquad 
d(v,u)=|v_0-u_0|+|v_1-u_1|+|v_2-u_2| 
\end{equation}
is a distance on $\mathcal{L}$, and
$u$ is a {\it neighbour of order} $l$ of $v$  if  $d(v,u)=l$.
The {\it nearest neighbours} of $v$ are
\begin{equation}
 \begin{array}{lll}
v^0=(v_0+\varepsilon (v),v_1,v_2) & & \\
v^1=(v_0,v_1+\varepsilon (v),v_2) & \qquad {\rm where} & \qquad 
\varepsilon (v)=(-1)^{v_0+v_1+v_2}\\
v^2=(v_0,v_1,v_2+\varepsilon (v)) & &
\end{array} 
\end{equation}
and the six {\it next-to-nearest neighbours} of $v$ are the points
$v^{ij}=(v^i)^j$ corresponding to all the pairs $(i,j)$ with $i\not=j.$
The {\it symmetry group} $G$ of the honeycomb lattice coincides with the group of all 
the isometries of the metric space  \ $(\mathcal{L},d),$ and is generated 
by the transformations 
\begin{equation}
 \begin{array}{l}
\mathcal{L}\longrightarrow \mathcal{L}:\quad (v_0,v_1,v_2)\mapsto (v_1,v_2,v_0)\\
\mathcal{L}\longrightarrow \mathcal{L}:\quad (v_0,v_1,v_2)\mapsto (v_0,v_2,v_1)\\
\mathcal{L}\longrightarrow \mathcal{L}:\quad (v_0,v_1,v_2)\mapsto (-v_0+1,-v_1,-v_2).
\end{array} 
\end{equation}
The subgroup of {\it translations} contained in $G$ corresponds to $\mathcal{T}$
\begin{equation}
\{ \ u\ |\  v\!\in \!\mathcal{L} \Rightarrow v\!+\!u\!\in \!\mathcal{L}\ \}
 =\{ \ u\in \mathcal{L}\ |\ \varepsilon (u)=1\ \}=\mathcal{T}.
\end{equation}

We can extend the description based on $e_0$, $e_1$, $e_2$ to the whole plane. 
Each vector $v$ admits the representation
\begin{equation}
 v=a\sum_{i=0}^2\langle v,e_i\rangle e_i\qquad {\rm with}\quad a=\frac{2}{3}
\end{equation}
and the usual scalar product and norm become
\begin{equation}
 \langle v,u\rangle =a\sum_{i=0}^2\langle v,e_i\rangle \langle u,e_i\rangle ,\qquad 
 ||v||=\sqrt{ a\sum_{i=0}^2\langle v,e_i\rangle ^2 }.
\end{equation}
The vectors $e_0$, $e_1$, $e_2$ form a tight frame \cite{An} and
a system of coherent vectors \cite{C1} in $\mathbb{R}^2$.
For each vector $v$, the `canonical coordinates'
\begin{equation}
 \hat v_0=\langle v,e_0\rangle,\qquad \hat v_1=\langle v,e_1\rangle ,\qquad 
   \hat v_2=\langle v,e_2\rangle 
\end{equation}
satisfy the relation $\hat v_0+\hat v_1+\hat v_2=0$, and we can identify $\mathbb{R}^2$ with the space
\begin{equation}
 \mathcal{K}=\{ k=(k_0,k_1,k_2)\ | \ k_0 ,\, k_1 ,\, k_2 \in \mathbb{R},\ k_0+k_1+k_2 =0\ \}
\end{equation} 
by using the linear isomorphism
\begin{equation}
\mathcal{K}\longrightarrow \mathbb{R}^2:\ (k_0,k_1,k_2)\mapsto a\sum_{i=0}^2k_ie_i. 
\end{equation}

The representation of a vector $v$ as a linear combination of $e_0,\, e_1,\, e_2$ is not unique
\begin{equation}
 v=a\sum_{i=0}^2\hat v_i\, e_i=a\sum_{i=0}^2(\hat v_i+\alpha )e_i
\end{equation} 
for any $\alpha \in \mathbb{R}$. All the elements $(v_0,\, v_1,\, v_2)$ of the set 
\[ \{ (\hat v_0+\alpha ,\, \hat v_1+\alpha ,\, \hat v_2+\alpha )\ |
\ \alpha \in \mathbb{R}\ \} \] 
correspond to the same point of the plane, and 
\begin{equation}
 \langle k,v\rangle =a\sum_{i=0}^2k_i\hat v_i=a\sum_{i=0}^2k_i(\hat v_i+\alpha )=a\sum_{i=0}^2k_i v_i
\end{equation}
for all $k\in \mathcal{K}$.

\section{An alternate mathematical model for carbon nanotubes}

\noindent A {\it single-wall carbon nanotube} can be visualized as the 
structure obtained by rolling a honeycomb lattice (which is a mathematical model 
for a graphene sheet) such that the endpoints
of a translation vector $c\in \mathcal{T}$ are folded one onto the other (see figure 1).
The vector $c$ is called the {\it chirality} of the tubule.
After the graphene sheet rolling, the points 
 $...,\ v-2c$,\ $v-c$,\ $v$,\  $v+c$,\ $v+2c,\ ... $
are folded one onto the other, for any $v\in \mathcal {L}$.
Therefore, each element of the set $[v]=v+\mathbb{Z}c$, that is, each element of the set
\begin{equation}
 [v_0,v_1,v_2]=\{ \ (v_0,v_1,v_2)+j(c_0,c_1,c_2)\ |\ j\in \mathbb{Z}\ \} 
\end{equation}
describe the same point of the nanotube.
The subset of the factor space $\mathbb{Z}^3/\mathbb{Z}c$ 
\begin{equation}
  \mathcal{L}_c=\left\{ \ \left. [v_0,v_1,v_2]\in \frac{\mathbb{Z}^3}{\mathbb{Z}c}\ 
\right| \ v_0+v_1+v_2\in \{ 0;1\} \ \right\} 
\end{equation}
can be regarded as a mathematical model for the nanotube of chirality $c$.
A symmetry transformation $\mathcal{L}\longrightarrow \mathcal{L}:\ v\mapsto gv$ 
of the honeycomb lattice satisfying the relation
\begin{equation}
 [v]=[u]\Longrightarrow [gv]=[gu] 
\end{equation}
defines a {\it symmetry transformation} of nanotube $\mathcal{L}_c$, namely 
\begin{equation}
 g:\mathcal{L}_c\longrightarrow \mathcal{L}_c:  [v]\mapsto [gv].
\end{equation} 
Each nanotube $\mathcal{L}_c$ admits the symmetry transformations:
\begin{equation} \begin{array}{l}
 \tau :\mathcal{L}_c\longrightarrow \mathcal{L}_c:  
[v_0,v_1,v_2]\mapsto [-v_0+1,-v_1,-v_2]\\ 
g_u:\mathcal{L}_c\longrightarrow \mathcal{L}_c:  [v]\mapsto [v+u]\qquad 
{\rm for\ any\qquad } u\in \mathcal{T}.
\end{array}
\end{equation} 

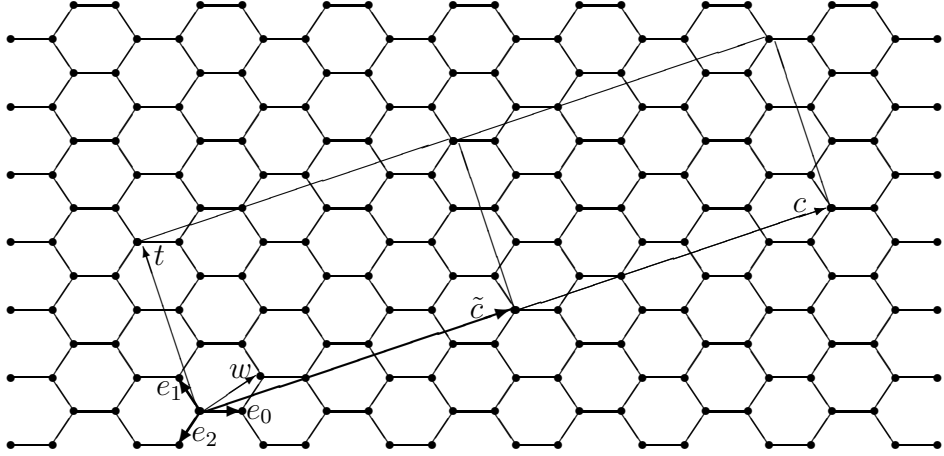
\begin{figure}[htbp]
\begin{center}
\setlength{\unitlength}{1mm}
\begin{picture}(120,60)
\multiput(0,0)(16.8,0){8}{\line(1,0){5.6}}
\multiput(16.8,0)(16.8,0){7}{\line(-2,3){3}}
\multiput(8.4,4.5)(16.8,0){7}{\line(-2,-3){3}}
\multiput(8.4,4.5)(16.8,0){7}{\line(1,0){5.6}}
\multiput(8.4,4.5)(16.8,0){7}{\line(-2,3){3}}
\multiput(16.8,9)(16.8,0){7}{\line(-2,-3){3}}
\multiput(0,9)(16.8,0){8}{\line(1,0){5.6}}
\multiput(16.8,9)(16.8,0){7}{\line(-2,3){3}}
\multiput(8.4,13.5)(16.8,0){7}{\line(-2,-3){3}}
\multiput(8.4,13.5)(16.8,0){7}{\line(1,0){5.6}}
\multiput(8.4,13.5)(16.8,0){7}{\line(-2,3){3}}
\multiput(16.8,18)(16.8,0){7}{\line(-2,-3){3}}
\multiput(0,18)(16.8,0){8}{\line(1,0){5.6}}
\multiput(16.8,18)(16.8,0){7}{\line(-2,3){3}}
\multiput(8.4,22.5)(16.8,0){7}{\line(-2,-3){3}}
\multiput(8.4,22.5)(16.8,0){7}{\line(1,0){5.6}}
\multiput(8.4,22.5)(16.8,0){7}{\line(-2,3){3}}
\multiput(16.8,27)(16.8,0){7}{\line(-2,-3){3}}
\multiput(0,27)(16.8,0){8}{\line(1,0){5.6}}
\multiput(16.8,27)(16.8,0){7}{\line(-2,3){3}}
\multiput(8.4,31.5)(16.8,0){7}{\line(-2,-3){3}}
\multiput(8.4,31.5)(16.8,0){7}{\line(1,0){5.6}}
\multiput(8.4,31.5)(16.8,0){7}{\line(-2,3){3}}
\multiput(16.8,36)(16.8,0){7}{\line(-2,-3){3}}
\multiput(0,36)(16.8,0){8}{\line(1,0){5.6}}
\multiput(16.8,36)(16.8,0){7}{\line(-2,3){3}}
\multiput(8.4,40.5)(16.8,0){7}{\line(-2,-3){3}}
\multiput(8.4,40.5)(16.8,0){7}{\line(1,0){5.6}}
\multiput(8.4,40.5)(16.8,0){7}{\line(-2,3){3}}
\multiput(16.8,45)(16.8,0){7}{\line(-2,-3){3}}
\multiput(0,45)(16.8,0){8}{\line(1,0){5.6}}
\multiput(16.8,45)(16.8,0){7}{\line(-2,3){3}}
\multiput(8.4,49.5)(16.8,0){7}{\line(-2,-3){3}}
\multiput(8.4,49.5)(16.8,0){7}{\line(1,0){5.6}}
\multiput(8.4,49.5)(16.8,0){7}{\line(-2,3){3}}
\multiput(16.8,54)(16.8,0){7}{\line(-2,-3){3}}
\multiput(0,54)(16.8,0){8}{\line(1,0){5.6}}
\multiput(16.8,54)(16.8,0){7}{\line(-2,3){3}}
\multiput(8.4,58.5)(16.8,0){7}{\line(-2,-3){3}}
\multiput(8.4,58.5)(16.8,0){7}{\line(1,0){5.6}}
\multiput(0,0)(16.8,0){8}{\circle*{1}}
\multiput(0,9)(16.8,0){2}{\circle*{1}}
\multiput(50.4,9)(16.8,0){5}{\circle*{1}}
\multiput(0,18)(16.8,0){8}{\circle*{1}}
\multiput(0,27)(16.8,0){8}{\circle*{1}}
\multiput(0,36)(16.8,0){8}{\circle*{1}}
\multiput(0,45)(16.8,0){8}{\circle*{1}}
\multiput(0,54)(16.8,0){8}{\circle*{1}}
\multiput(5.6,0)(16.8,0){8}{\circle*{1}}
\multiput(5.6,9)(16.8,0){8}{\circle*{1}}
\multiput(5.6,18)(16.8,0){8}{\circle*{1}}
\multiput(5.6,27)(16.8,0){8}{\circle*{1}}
\multiput(5.6,36)(16.8,0){8}{\circle*{1}}
\multiput(5.6,45)(16.8,0){8}{\circle*{1}}
\multiput(5.6,54)(16.8,0){8}{\circle*{1}}
\multiput(8.4,4.5)(16.8,0){7}{\circle*{1}}
\multiput(8.4,13.5)(16.8,0){7}{\circle*{1}}
\multiput(8.4,22.5)(16.8,0){7}{\circle*{1}}
\multiput(8.4,31.5)(16.8,0){7}{\circle*{1}}
\multiput(8.4,40.5)(16.8,0){7}{\circle*{1}}
\multiput(8.4,49.5)(16.8,0){7}{\circle*{1}}
\multiput(8.4,58.5)(16.8,0){7}{\circle*{1}}
\multiput(14,4.5)(16.8,0){7}{\circle*{1}}
\multiput(14,13.5)(16.8,0){7}{\circle*{1}}
\multiput(14,22.5)(16.8,0){7}{\circle*{1}}
\multiput(14,31.5)(16.8,0){7}{\circle*{1}}
\multiput(14,40.5)(16.8,0){7}{\circle*{1}}
\multiput(14,49.5)(16.8,0){7}{\circle*{1}}
\multiput(14,58.5)(16.8,0){7}{\circle*{1}}
\put(25.2,4.3){\vector(3,2){7.4}}
\put(24.9,4.5){\vector(-1,3){7.3}}
\put(67,17.8){\vector(3,1){41.3}}
\put(16.9,27){\line(3,1){41.3}}
\put(67,18){\line(-1,3){7.4}}
\put(67,17.8){\vector(3,1){41.3}}
\put(58.9,40.5){\line(3,1){41.3}}
\put(109,31.5){\line(-1,3){7.4}}
\put(25,4.5){\circle*{1}}
\put(33.2,9.2){\circle*{1}}
\put(16.9,27){\circle*{1}}
\put(67,18){\circle*{1}}
\put(109,31.5){\circle*{1}}
\put(58.9,40.5){\circle*{1}}
\put(24,0.5){$e_2$}
\put(19.3,6.7){$e_1$}
\put(31.3,3.5){$e_0$}
\put(104,31){$c$}
\put(61,17){$\tilde{c}$}
\put(29,9){$w$}
\put(19,24){$t$}
\thicklines
\put(25.2,4.5){\vector(1,0){5.6}}
\put(25.2,4.5){\vector(-2,3){3}}
\put(25.2,4.5){\vector(-2,-3){3}}
\put(25.2,4.3){\vector(3,1){41.3}}
\end{picture}
\end{center}
\caption{The unit cell of the carbon nanotube with the chiral vector $c=(10,-2,-8)$.
In this case $n=2$, $\tilde{c}=(5,-1,-4)$, $t=(-1,3,-2)$, $w=(1,0,-1)$ and $\tilde q=14.$}
\end{figure}

Let \ $n$ be the greatest common divisor of $c_0$, $c_1$, $c_2$, and let
$\tilde c=(\tilde c_0,\tilde c_1,\tilde c_2)$, where
\begin{equation}
\tilde c_0=\frac{1}{n}c_0,\qquad  \tilde c_1=\frac{1}{n}c_1,\qquad  
\tilde c_2=\frac{1}{n}c_2.
\end{equation}
The transformation \ $g_{\tilde c}$ \ represents a {\it rotation of nanotube} of 
angle $2\pi /n$ with respect to its axis.
Since 
\begin{equation}
(c_2-c_1)c_0+(c_0-c_2)c_1+(c_1-c_0)c_2=0
\end{equation}
 the vector $(c_2-c_1,\, c_0-c_2,\, c_1-c_0)$ is orthogonal to $c$ (it has the direction
of nanotube axis).
The transformation \ $g_t$ \ corresponding to
\begin{equation}
t=\frac{1}{\mathcal{R}}(c_2-c_1,\, c_0-c_2,\, c_1-c_0)
\end{equation}
where
\begin{equation}
\fl  \mathcal{R}={\rm gcd}\{c_2-c_1,\, c_0-c_2,\, c_1-c_0\}=\left\{ \begin{array}{rll}
3n & \quad {\rm if} & \quad \tilde c_2-\tilde c_1 \in 3\mathbb{Z}\\
n & \quad {\rm if} & \quad \tilde c_2-\tilde c_1 \not\in 3\mathbb{Z}
\end{array} \right. 
\end{equation}
represents the shortest pure translation of nanotube.
From $c_0+c_1+c_2=0$ we get 
\begin{equation}
(c_1-c_2)^2+(c_2-c_0)^2+(c_0-c_1)^2=3(c_0^2+c_1^2+c_2^2)
\end{equation} 
that is, $\mathcal{R}^2||t||^2=3||c||^2$, whence
\begin{equation}
q=\frac{1}{\mathcal {R}}(c_0^2+c_1^2+c_2^2)\in n\mathbb{Z}.
\end{equation}
For any $u\in \mathcal {T}$ the projections of $u$ on $c$ and $t$
can be written 
\begin{equation}
\fl \frac{\langle u,c\rangle }{||c||^2}c
=\left( u_1\frac{c_1-c_0}{\mathcal{R}}+u_2\frac{c_2-c_0}{\mathcal{R}}\right)
\frac{c}{q}\, ,\qquad 
\frac{\langle u,t\rangle }{||t||^2}t
=\left( u_1\frac{c_2}{n}-u_2\frac{c_1}{n}\right)\frac{t}{\tilde q}
\end{equation}
where   $\tilde q=\frac{q}{n}.$
Since 
\begin{equation}
 \fl {\rm gcd}\{(c_1-c_0)/\mathcal{R},(c_2-c_0)/\mathcal{R}\}=1\qquad  {\rm and}\qquad 
{\rm gcd}\{ c_2/n,c_1/n\}=1
\end{equation} 
it follows that the projection of $\mathcal{T}$ on $c$ is $\mathbb{Z}\frac{c}{q}$,
and the projection of $\mathcal{T}$ on $t$ is $\mathbb{Z}\frac{t}{\tilde q}$. 
Let $w \in \mathcal{T}$ be the shortest vector with
\begin{equation} 
\frac{\langle w ,t\rangle }{||t||^2}t=\frac{t}{\tilde q}.
\end{equation}
From the relation 
\begin{equation}
w=\frac{\langle w,c\rangle }{||c||^2}c+\frac{\langle w,t\rangle }{||t||^2}t
\end{equation}
we get $[qw]=[nt].$

Without lost of generality we can assume $c_0>c_1\geq c_2$.
In the case $c_1=c_2$ we have an {\it armchair} nanotube, and in the case
$c_1=0$ a {\it zig-zag} nanotube. The nanotubes with $0\not=c_1\not=c_2$ are 
called {\it chiral nanotubes}. Our approach works for any single-wall 
carbon nanotube, but in this paper we restrict us to chiral nanotubes.
The symmetry group  \ $G_c$ of the nanotube $\mathcal{L}_c$ 
is generated by the transformations  $\varrho =g_{\tilde c}$, $\sigma =g_w$ and $\tau $
\begin{equation}
G_c=\left\langle \varrho ,\, \sigma ,\, \tau \ \ \left| \ \ \varrho \sigma =\sigma \varrho ,\ \ 
\varrho ^n=\tau ^2=(\sigma \tau )^2=(\varrho \tau )^2=e\ \right. \right\rangle  .
\end{equation}
For each \ $[v]\in \mathcal{L}_c$ \ 
there exist $s\in \mathbb{Z}$, $m\in \{ 0,\, 1,\, ...,\, n-1\}$ and $p\in \{ 0; 1\}$ 
uniquely determined such that \ \ $ [v]=\sigma ^s\, \varrho ^m\, \tau ^p[0,0,0].$ 
The usual description \cite{D0,Vu} of the atomic positions of the atoms forming a carbon nanotube
is based on this remark, and the set 
\begin{equation}
 \{ (s,m,p)\ |\ s\in \mathbb{Z},\ m\in \{ 0,1,...,n-1\},\ p\in \{ 0,1\}\}
\end{equation}
is used as a mathematical model. The alternate mathematical model we present
in this paper is $\mathcal{L}_c$.
\section{A tight-binding approach to carbon nanotubes}
\noindent Consider the Hilbert space
$(l^2({\mathcal L}_c),\langle , \rangle )$, where
\begin{equation} 
l^2({\mathcal L}_c)=\left\{ \psi :{\mathcal L}_c\longrightarrow \mathbb{C}\left| 
\ \ \sum_{v\in {\mathcal L}_c}|\psi (v)|^2<\infty \right. \right\} 
\end{equation}
\begin{equation}
\langle \psi _1,\psi _2\rangle =
\sum_{v\in {\mathcal L}_c}\overline{\psi _1(v)}\, \psi _2(v)
\end{equation}
and the unitary representation of $G_c$ in $l^2({\mathcal L}_c)$ defined by 
\begin{equation}
g:l^2({\mathcal L}_c)\longrightarrow l^2({\mathcal L}_c)\qquad 
(g\psi )[v]=\psi (g^{-1}[v]).
\end{equation}
For each $\kappa \in (0,\infty )$, the linear operator
\begin{equation} \label{Hamiltonian}
H:l^2({\mathcal L}_c)\longrightarrow l^2({\mathcal L}_c)\qquad
(H\psi )[v]=\kappa \sum_{j=0}^2 \psi [v^j]
\end{equation}
is a self-adjoint $G_c$-invariant operator.

If $k\in \mathcal{K}$ is such that \ $\langle k,c\rangle \in 2\pi \mathbb{Z}$ then
\begin{equation}
{\rm e}^{{\rm i}\langle k,v\rangle }={\rm e}^{{\rm i}\langle k,v+jc\rangle }
\qquad {\rm for\ any}\ \ j\in \mathbb{Z}
\end{equation}
and hence, the function 
\begin{equation}
\mathcal{L}_c\longrightarrow \mathbb{C}:\ 
[v]\mapsto {\rm e}^{-{\rm i}\langle k,v\rangle }
\end{equation}
is well-defined (it does not depend on the representative we choose for $[v]$).
The Hamiltonian used in the 
tight-binding description of $\pi $ bands has the form (\ref{Hamiltonian}).\\[5mm]
{\bf Theorem 1.} {\it 
a) For any $k$ \ such that $\langle k,c\rangle \!\!\in \!\!2\pi \mathbb{Z}$
the numbers $\pm E(k)$, \ where
\begin{equation} 
\fl \begin{array}{l}
 E(k)=\kappa | {\rm e}^{ {\rm i}k_0a}+ {\rm e}^{ {\rm i}k_1a}
+ {\rm e}^{ {\rm i}k_2a}|\\[2mm] 
\mbox{}\qquad \, =\kappa  \sqrt{3+2\cos(k_0-k_1)a+2\cos(k_1-k_2)a+2\cos(k_2-k_0)a}
\end{array}
\end{equation} 
belong to the spectrum of $H$.\\
b) The bounded functions belonging to an extension of $l^2(\mathcal{L}_c)$  
\begin{equation}
\psi _k^\pm :\mathcal{L}_c\longrightarrow \mathbb{C},\qquad 
\psi _k^\pm [v]={\rm e}^{-{\rm i}\langle k,v\rangle }\varphi _k^\pm [v] 
\end{equation} 
where
\begin{equation}
\varphi _k^\pm [v]=
\left\{ \begin{array}{rll}
{\rm e}^{{\rm i}\lambda (k)}
& \ \ if  & \ \ \varepsilon (v)=1\\
\pm {\rm e}^{-{\rm i}\lambda (k)}
 &\ \  if & \ \ \varepsilon (v)=-1
\end{array} \right. 
\end{equation}
and
\begin{equation} \fl 
 \lambda (k)=\left\{ \begin{array}{lll}
-\frac{1}{2}\, {\rm arg}
\left({\rm e}^{{\rm i}k_0a}+{\rm e}^{{\rm i}k_1a}+{\rm e}^{{\rm i}k_2a}\right) & if &
{\rm e}^{{\rm i}k_0a}+{\rm e}^{{\rm i}k_1a}+{\rm e}^{{\rm i}k_2a}\not= 0\\[2mm]
0 & if &
{\rm e}^{{\rm i}k_0a}+{\rm e}^{{\rm i}k_1a}+{\rm e}^{{\rm i}k_2a}= 0
\end{array} \right. 
\end{equation}
are eigenfunctions of $H$ corresponding to the eigenvalues $\pm E(k)$, that is, 
\begin{equation}
H\, \psi _k^\pm =\pm E(k)\, \psi _k^\pm . 
\end{equation} }
\noindent {\bf Proof.}
a) The function $\psi :\mathcal{L}_c\longrightarrow \mathbb{C}$,   
$\psi [v]={\rm e}^{-{\rm i}\langle k,v\rangle }\varphi [v]$, where
\begin{equation}
\varphi [v]=\left\{ \begin{array}{lll}
\alpha \ \ & {\rm if} & \varepsilon (v)=1\\
\beta & {\rm if } & \varepsilon (v)=-1 
\end{array}\right. 
\end{equation}
and $\alpha $, $\beta $ are two constants, satisfies the relation 
$H\psi =E\psi $ if and only if $(\alpha ,\beta )$ is a solution of the 
system of equations
\begin{equation} 
\left\{ \begin{array}{r}
\kappa ({\rm e}^{ -{\rm i}k_0a}+{\rm e}^{ -{\rm i}k_1a}+
{\rm e}^{ -{\rm i}k_2a})\beta =E\alpha \ \mbox{}\\
\kappa ({\rm e}^{{\rm i}k_0a}+ {\rm e}^{{\rm i}k_1a}+{\rm e}^{{\rm i}k_2a})\alpha =E\beta .
\end{array} \right. 
\end{equation}
This system has non-trivial solutions if and only if
\begin{equation}
\left| \begin{array}{cc}
-E & \kappa ({\rm e}^{ -{\rm i}k_0a}+{\rm e}^{ -{\rm i}k_1a}+{\rm e}^{ -{\rm i}k_2a})\\
\kappa ({\rm e}^{{\rm i}k_0a}+{\rm e}^{{\rm i}k_1a}+{\rm e}^{{\rm i}k_2a}) & -E
\end{array} \right| =0
\end{equation}
that is, if and only if $E$ is one of the numbers $\pm E(k) $.\\
b) If ${\rm e}^{{\rm i}k_0a}+{\rm e}^{{\rm i}k_1a}+{\rm e}^{{\rm i}k_0a}\not= 0$
then the equation 
$\kappa ({\rm e}^{{\rm i}k_0a}+ {\rm e}^{{\rm i}k_1a}+{\rm e}^{{\rm i}k_2a})\alpha =\pm E(k)\beta $
leads to
\begin{equation}
\beta = \pm \alpha \frac{ {\rm e}^{ {\rm i}k_0a}+ {\rm e}^{ {\rm i}k_1a}+ {\rm e}^{ {\rm i}k_2a}}
{| {\rm e}^{ {\rm i}k_0a}+ {\rm e}^{ {\rm i}k_1a}+ {\rm e}^{ {\rm i}k_2a}|}=
 \pm \alpha \, {\rm e}^{-2{\rm i}\lambda (k)} .
\end{equation}
Choosing $\alpha ={\rm e}^{{\rm i}\lambda (k)}$ we get $\beta =\pm {\rm e}^{-{\rm i}\lambda (k)}$, and hence,
up to a multiplicative constant, the solution 
of $H\, \psi =\pm E(k)\, \psi $ is $\psi =\psi _k^\pm $.
If ${\rm e}^{{\rm i}k_0a}+{\rm e}^{{\rm i}k_1a}+{\rm e}^{{\rm i}k_0a}= 0$ then $E(k)=0$, and the functions
\begin{equation}
 \psi _k^\pm [v]={\rm e}^{-{\rm i}\langle k,v\rangle }\left\{ \begin{array}{lll}
1 \ \ & {\rm if} & \varepsilon (v)=1\\
\pm 1 & {\rm if } & \varepsilon (v)=-1 
\end{array}\right. 
\end{equation}
are eigenfunctions.\qquad $\rule{1mm}{1mm}$

\noindent The relation $\langle k,c\rangle \in 2\pi \mathbb{Z}$
defines a family of equidistant straight lines orthogonal to \, $c$
with the distance between neighbouring lines equal to $2\pi /||c||$.\\[5mm]
{\bf Theorem 2.} {\it 
a) The function 
\begin{equation}
E:\mathcal{K}\longrightarrow [0,3\kappa ]\qquad    
E(k)=\kappa | {\rm e}^{ {\rm i}k_0a}+ {\rm e}^{ {\rm i}k_1a}
+ {\rm e}^{ {\rm i}k_2a}|
\end{equation} 
is even and periodic 
\begin{equation}
E(k)\!=\!E(-k)\qquad \quad E(k)\!=\!E(k\!+\!b_0)\!=\!E(k\!+\!b_1)\!=\!E(k\!+\!b_2)
\end{equation}
where
\begin{equation}
\fl b_0=\left( \frac{4\pi }{3a},-\frac{2\pi }{3a},-\frac{2\pi }{3a}\right),\quad 
b_1=\left( -\frac{2\pi }{3a},\frac{4\pi }{3a},-\frac{2\pi }{3a}\right),\quad 
b_2=\left( -\frac{2\pi }{3a},-\frac{2\pi }{3a},\frac{4\pi }{3a}\right). 
\end{equation}
b) The functions \ $\psi _k=\psi _k^+$ and $\psi _k^-$\ 
are eigenfunctions of \ $\varrho $ \ and \ $\sigma $ 
\begin{equation}
\varrho \, \psi _k^\pm 
={\rm e}^{{\rm i}\langle k,\tilde c\rangle }\, \psi _k^\pm \qquad \quad 
 \sigma \, \psi _k^\pm 
={\rm e}^{{\rm i}\langle k,w\rangle }\, \psi _k^\pm 
\end{equation}
satisfy the relation
\begin{equation}
\tau \, \psi _k^\pm \!=\!\pm {\rm e}^{-{\rm i}k_0a }\, \psi _{-k}^\pm .
\end{equation}
c) The eigenspaces 
\begin{equation} 
\begin{array}{l}
\mathcal{E}_k \ =\{ \ \alpha \psi _k +\beta \psi _{-k} \ \ |\ \ 
\alpha,\, \beta \in \mathbb{C}\ \ \} \\
\mathcal{E}_k^- =\{ \ \alpha \psi _k^- +\beta \psi _{-k}^- \ \ |\ \ 
\alpha,\, \beta \in \mathbb{C}\ \ \}
\end{array}
\end{equation}
corresponding to $E(k)$ and $-E(k)$ are $G_c$-invariant, 
$\mathcal{E}_k =\mathcal{E}_{-k} $\ and \ $\mathcal{E}_k^- =\mathcal{E}_{-k}^- $.\\
d) If $k$ is such that $E(k)\not=0$ then 
\begin{equation}
\psi _{k+b_i}^\pm ={\rm e}^{{\rm i}\pi /3}\psi _k^\pm \qquad \quad 
\mathcal{E}_{k+b_i}=\mathcal{E}_k 
\end{equation}
for any}  $i\in \{ 0,1,2\}$.\\[5mm]

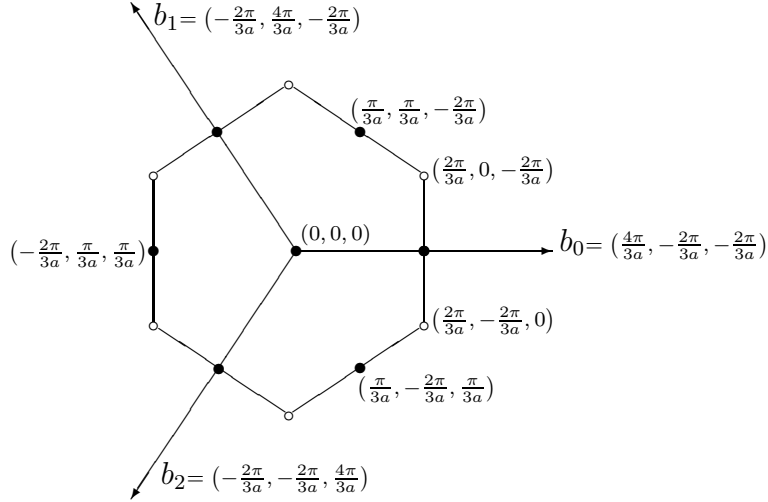
\begin{figure}
\begin{center}
\setlength{\unitlength}{1mm}
\begin{picture}(100,70)
\put(40.5,10.5){\line(3,2){16.8}}
\put(39.5,10.5){\line(-3,2){16.8}}
\put(58,22.6){\line(0,1){18.8}}
\put(22,22.6){\line(0,1){18.8}}
\put(22.5,42.5){\line(3,2){16.8}}
\put(57.5,42.5){\line(-3,2){16.8}}
\put(41,32){\vector(1,0){34}}
\put(41,32){\vector(-2,3){22}}
\put(41,32){\vector(-2,-3){22}}
\put(41,32){\circle*{1.5}}
\put(58,32){\circle*{1.5}}
\put(58,32){\circle*{1.5}}
\put(22,32){\circle*{1.5}}
\put(30.7,16.3){\circle*{1.5}}
\put(30.5,47.8){\circle*{1.5}}
\put(49.5,16.4){\circle*{1.5}}
\put(49.5,47.8){\circle*{1.5}}
\put(58,22){\circle{1}}
\put(22,22){\circle{1}}
\put(58,42){\circle{1}}
\put(22,42){\circle{1}}
\put(40,54){\circle{1}}
\put(40,10){\circle{1}}
\put(76,32){$b_0${\scriptsize $=\left( \frac{4\pi }{3a},-\frac{2\pi }{3a},-\frac{2\pi }{3a}\right)$}}
\put(23,1){$b_2${\scriptsize $=\left( -\frac{2\pi }{3a},-\frac{2\pi }{3a},\frac{4\pi }{3a}\right)$}}
\put(22,62){$b_1${\scriptsize $=\left( -\frac{2\pi }{3a},\frac{4\pi }{3a},-\frac{2\pi }{3a}\right)$}}
\put(2.7,31){{\scriptsize $\left( -\frac{2\pi }{3a},\frac{\pi }{3a},\frac{\pi }{3a}\right)$}}
\put(48,50){{\scriptsize $\left( \frac{\pi }{3a},\frac{\pi }{3a},-\frac{2\pi }{3a}\right)$}}
\put(49,13){{\scriptsize $\left( \frac{\pi }{3a},-\frac{2\pi }{3a},\frac{\pi }{3a}\right)$}}
\put(41.5,33.5){{\scriptsize $\left( 0,0,0\right)$}}
\put(58.8,42){{\scriptsize $\left( \frac{2\pi }{3a},0,-\frac{2\pi }{3a}\right)$}}
\put(58.8,22){{\scriptsize $\left( \frac{2\pi }{3a},-\frac{2\pi }{3a},0\right)$}}
\end{picture}
\end{center}
\caption{ The hexagonal domain $\mathcal{B}$, the points of $\Lambda $ (indicated by $\bullet $) 
and the points $k$ with $E(k)=0$, called K points (indicated by $\circ $).}
\end{figure}

\noindent {\bf Proof.}  Since ${\rm e}^{{\rm i}4\pi /3}={\rm e}^{-{\rm i}2\pi /3}$ we have
\[
E(k+b_0)=\kappa \left| {\rm e}^{{\rm i}k_0a}\, {\rm e}^{{\rm i}4\pi /3}+
{\rm e}^{{\rm i}k_1a}\, {\rm e}^{-{\rm i}2\pi /3}
+{\rm e}^{{\rm i}k_2a}\, {\rm e}^{-{\rm i}2\pi /3}\right| =E(k).
\] 
\[
\lambda (k+b_0)=-\frac{1}{2}\, {\rm arg}
\left[\left({\rm e}^{{\rm i}k_0a}+{\rm e}^{{\rm i}k_1a}+{\rm e}^{{\rm i}k_2a}\right) 
{\rm e}^{-{\rm i}2\pi /3}\right]=\lambda (k)+\frac{\pi }{3} 
\]
for any $k$ with $E(k)\not=0$. 
If $v$ is such that $\varepsilon (v)=1$ then $\langle b_0,v\rangle =2\pi v_0$ and
\[ \fl
\psi _{k+b_0}[v]={\rm e}^{-{\rm i}\langle k+b_0,v\rangle }\, {\rm e}^{{\rm i}\lambda (k+b_0)}
={\rm e}^{-{\rm i}\langle k,v\rangle }\, {\rm e}^{-{\rm i}2\pi v_0 }\, {\rm e}^{{\rm i}\lambda (k)}\, 
{\rm e}^{{\rm i}\pi /3}={\rm e}^{{\rm i}\pi /3}\,  \psi _k[v].
\]
If $v$ is such that $\varepsilon (v)=-1$ then $\langle b_0,v\rangle =2\pi v_0-2\pi /3$ and
\[ \fl
\psi _{k+b_0}[v]={\rm e}^{-{\rm i}\langle k+b_0,v\rangle }\, {\rm e}^{-{\rm i}\lambda (k+b_0)}
={\rm e}^{-{\rm i}\langle k,v\rangle }\, {\rm e}^{-{\rm i}2\pi v_0 }\, {\rm e}^{{\rm i}2\pi /3}
{\rm e}^{-{\rm i}\lambda (k)}\, {\rm e}^{-{\rm i}\pi /3}={\rm e}^{{\rm i}\pi /3}\,  \psi _k[v].
\qquad \rule{1mm}{1mm} \]

\noindent From the periodicity of $E(k)$ and $\mathcal{E}_k$ it follows that we can restrict our analysis to
the case $k\in \mathcal{B}_c$, where
\begin{equation}
\mathcal{B}_c=\{ \ k\in \mathcal{B}\ |\ \ \langle k,c\rangle \in 2\pi \mathbb{Z} \ \}
\end{equation}
and $\mathcal{B}$ is the hexagonal domain (see figure 2)
\begin{equation} 
\mathcal{B}=\left\{ k\in \mathcal{K} \left| \ -\frac{2\pi }{3a}\leq k_i\leq \frac{2\pi }{3a}\ \ 
{\rm for \ any\ } \ i\in \{ 0,1,2\} \right. \right\}.
\end{equation}

\noindent {\bf Theorem 3.} {\it For even $n$ the space 
$\mathcal{E}_k$ is one-dimensional if and only if $k\in \Lambda $, where}
\[ \fl 
\Lambda =\left\{ (0,0,0),\, \pm \left( -\frac{2\pi }{3a},\frac{\pi }{3a},\frac{\pi }{3a}\right),\, 
\pm \left( \frac{\pi }{3a},-\frac{2\pi }{3a},\frac{\pi }{3a}\right),\, 
\pm \left( \frac{\pi }{3a},\frac{\pi }{3a},-\frac{2\pi }{3a}\right)\right\}. 
\]
{\bf Proof.} If $\mathcal{E}_k$ is one-dimensional then there is a constant $C$ such that 
\[ \psi _{-k}[v]=C\psi _k[v]\qquad {\rm for\ any}\ \ v\in \mathcal{L}_c .\]
For $v=(0,0,0)$ we get $C={\rm e}^{-2{\rm i}\lambda (k)}$. Therefore
\begin{equation}
\psi _{-k}=C\, \psi _k \quad \Longleftrightarrow \quad 
\left\{ \begin{array}{lll}
{\rm e}^{2{\rm i}\langle k,v\rangle }=1 & {\rm if} & \varepsilon (v)=1\\
{\rm e}^{2{\rm i}\langle k,v\rangle }=C^2 & {\rm if} & \varepsilon (v)=-1.
\end{array} \right. 
\end{equation}
Particularly, we must have
${\rm e}^{2{\rm i}k_0a}={\rm e}^{2{\rm i}k_1a}={\rm e}^{2{\rm i}k_2a}$. 
For $k\in \mathcal{B}$ this relation is possible only if there are 
$\alpha , \beta \in \{ 0,\pm 1\}$ such that
$k_1=k_0+\alpha \pi /a$ and $k_2=k_0+\beta \pi /a$.
From $k_0+k_1+k_2=0$ we get
\begin{equation}
 k=\left( -\frac{\alpha +\beta }{3}\frac{\pi }{a},
\frac{2\alpha -\beta }{3}\frac{\pi }{a},\frac{-\alpha +2\beta }{3}\frac{\pi }{a}\right).
\end{equation}
All these points points lie on the family of straight line 
defined by $\langle k,c\rangle \in 2\pi \mathbb{Z}$ since
\begin{equation} \label{Lc}
 \fl \langle k,c\rangle =k_0c_0a+\left( k_0+\alpha \frac{\pi }{a}\right)c_1a
+\left( k_0+\beta \frac{\pi }{a}\right)c_2a=(\alpha c_1+\beta c_2)\pi \in 2\pi \mathbb{Z}.
\end{equation}
In the case $\alpha =\beta =0$ we get $k=(0,0,0)$, $\psi _k[v]=1$, and
\begin{equation}
\varrho \psi _k=\psi _k\qquad \sigma \psi _k=\psi _k\qquad \tau \psi _k=\psi _k. 
\end{equation}
In the case $\alpha =\beta =1$ we get $k=(-2\pi /3a, \pi /3a,\pi /3a)$, 
$\psi _k[v]=(-1)^{v_0}{\rm e}^{-{\rm i}\pi /6}$, and
\begin{equation}
 \varrho \psi _k=(-1)^{\tilde{c}_0}\psi _k\qquad 
\sigma \psi _k=(-1)^{w_0}\psi _k\qquad \tau \psi _k=-\psi _k .
\end{equation}
All the other points $k\in \mathcal{B}_c$ with ${\rm dim}\, \mathcal{E}_k=1$  can be obtained 
from the analyzed points by using permutations of coordinates and/or a multiplication by 
(-1).\qquad $\rule{1mm}{1mm}$\\[2mm]
In the case when $n$ is odd only a part of the points of $\Lambda $ lies on the 
family of straight lines defined by $\langle k,c\rangle \in 2\pi \mathbb{Z}$.

\section{Two-dimensional representations of $G_c$}

If $k\in \mathcal{B}_c\backslash \Lambda $ then ${\rm dim}\, \mathcal{E}_k=2$ and the matrices of 
$\varrho $, $\sigma $ and $\tau $ in the basis $\{ \psi _k,\, \psi _{-k}\}$ are
\begin{equation} \label{rep2}
\fl \varrho =\left( \begin{array}{ll}
{\rm e}^{{\rm i}\langle k,\tilde c\rangle } & 0\\
0 & {\rm e}^{-{\rm i}\langle k,\tilde c\rangle }
\end{array} \right), \quad 
\sigma =\left( \begin{array}{ll}
{\rm e}^{{\rm i}\langle k,w\rangle } & 0\\
0 & {\rm e}^{-{\rm i}\langle k,w\rangle }
\end{array} \right), \quad 
\tau =\left( \begin{array}{ll}
0 & {\rm e}^{{\rm i}k_0a} \\
{\rm e}^{-{\rm i}k_0a } & 0
\end{array} \right) 
\end{equation}  
{\bf Theorem 4.} {\it For each $k\in \mathcal{B}_c\backslash \Lambda $ 
the representations of $G_c$ in $\mathcal{E}_k$ and $\mathcal{E}_k^-$ are equivalent.
They are reducible if and only if the numbers 
$\langle k,\tilde c\rangle $ and $\langle k,w\rangle $ belong to $\mathbb{Z}\pi $.}\\[3mm]
{\bf Proof.} The linear transformation 
$\mathcal{E}_k\longrightarrow \mathcal{E}_k^-:\alpha \psi _k\!+\!\beta \psi _{-k} \mapsto
{\rm i}\alpha \psi _k^-\!-\!{\rm i}\beta \psi _{-k}^- $
is an isomorphism of representations. If 
\[ 
P=\left( \begin{array}{ll}
\alpha & \beta \\
\gamma & \delta 
\end{array} \right) \]
is the matrix in the basis $\{ \psi _k,\, \psi _{-k}\}$ of the projector corresponding to 
a $G_c$-invariant subspace of $\mathcal{E}_k$ then  $P^2=P$ and
\begin{equation}
\left( \begin{array}{ll}
\alpha & \beta \\
\gamma & \delta 
\end{array} \right) 
\left( \begin{array}{ll}
{\rm e}^{{\rm i}\langle k,\tilde c\rangle } & 0\\
0 & {\rm e}^{-{\rm i}\langle k,\tilde c\rangle }
\end{array} \right)=
\left( \begin{array}{ll}
{\rm e}^{{\rm i}\langle k,\tilde c\rangle } & 0\\
0 & {\rm e}^{-{\rm i}\langle k,\tilde c\rangle }
\end{array} \right)\left( \begin{array}{ll}
\alpha & \beta \\
\gamma & \delta 
\end{array} \right)
\end{equation}  
\begin{equation}
\left( \begin{array}{ll}
\alpha & \beta \\
\gamma & \delta 
\end{array} \right) 
\left( \begin{array}{ll}
{\rm e}^{{\rm i}\langle k,w\rangle } & 0\\
0 & {\rm e}^{-{\rm i}\langle k,w\rangle }
\end{array} \right)
=\left( \begin{array}{ll}
{\rm e}^{{\rm i}\langle k,w\rangle } & 0\\
0 & {\rm e}^{-{\rm i}\langle k,w\rangle }
\end{array} \right) 
\left( \begin{array}{ll}
\alpha & \beta \\
\gamma & \delta 
\end{array} \right)
\end{equation}  
\begin{equation}
\left( \begin{array}{ll}
\alpha & \beta \\
\gamma & \delta 
\end{array} \right) 
\left( \begin{array}{ll}
0 & {\rm e}^{{\rm i}k_0a} \\
{\rm e}^{-{\rm i}k_0a } & 0
\end{array} \right)
=\left( \begin{array}{ll}
0 & {\rm e}^{{\rm i}k_0a} \\
{\rm e}^{-{\rm i}k_0a } & 0
\end{array} \right)   
\left( \begin{array}{ll}
\alpha & \beta \\
\gamma & \delta 
\end{array} \right).
\end{equation}  
From the last three relations it follows that
\begin{equation}
 \beta {\rm e}^{{\rm i}\langle k,\tilde c\rangle } =\beta {\rm e}^{-{\rm i}\langle k,\tilde c\rangle } \qquad 
\beta {\rm e}^{{\rm i}\langle k,w\rangle } =\beta {\rm e}^{-{\rm i}\langle k,w\rangle }\qquad  
\alpha {\rm e}^{{\rm i}k_0a }=\delta {\rm e}^{{\rm i}k_0a } 
\end{equation}
\begin{equation}
\gamma {\rm e}^{{\rm i}\langle k,\tilde c\rangle } =\gamma {\rm e}^{-{\rm i}\langle k,\tilde c\rangle } \qquad 
\gamma  {\rm e}^{{\rm i}\langle k,w\rangle } =\gamma {\rm e}^{-{\rm i}\langle k,w\rangle } \qquad 
\beta {\rm e}^{-{\rm i}k_0a }=\gamma {\rm e}^{{\rm i}k_0a } .
\end{equation}
Since ${\rm e}^{{\rm i}k_0a }\not =0$ we obtain $\alpha =\delta $.
If either $\langle k,\tilde c\rangle \not \in \mathbb{Z}\pi $ or
$\langle k,w\rangle \not \in \mathbb{Z}\pi $ then
$\beta =\gamma =0$ and the representation (\ref{rep2}) is irreducible. 
If the numbers $\langle k,\tilde c\rangle $ and $\langle k,w\rangle $ belong to $\mathbb{Z}\pi $ 
then imposing the condition $P^2=P$ we obtain 
\begin{equation}
P=\frac{1}{2}\left( \begin{array}{cc}
1 & {\rm e}^{{\rm i}k_0a } \\[2mm]
{\rm e}^{-{\rm i}k_0a }  & 1
\end{array} \right) \qquad {\rm or} \qquad 
P=\frac{1}{2}\left( \begin{array}{cc}
1 & -{\rm e}^{{\rm i}k_0a } \\[2mm]
-{\rm e}^{-{\rm i}k_0a }  & 1
\end{array} \right) .
\end{equation}
These complementary projectors correspond to the decomposition 
\begin{equation}
 \mathcal{E}_k=\{ \alpha (\psi _k+{\rm e}^{-{\rm i}k_0a }\psi _{-k})\ |\ \alpha \in \mathbb{C}\ \} \oplus  
\{ \alpha (\psi _k-{\rm e}^{-{\rm i}k_0a }\psi _{-k})\ |\ \alpha \in \mathbb{C}\ \} 
\end{equation}
of $\mathcal{E}_k$ into direct sum of one-dimensional $G_c$-invariant subspaces, and
\[
\begin{array}{l}
\varrho (\psi _k\pm {\rm e}^{-{\rm i}k_0a }\psi _{-k})
=(-1)^m (\psi _k\pm {\rm e}^{-{\rm i}k_0a }\psi _{-k})\\[2mm]
\sigma (\psi _k\pm {\rm e}^{-{\rm i}k_0a }\psi _{-k})
=(-1)^p (\psi _k\pm {\rm e}^{-{\rm i}k_0a }\psi _{-k})\\[2mm]
\tau (\psi _k\pm {\rm e}^{-{\rm i}k_0a }\psi _{-k})
=\pm (\psi _k\pm {\rm e}^{-{\rm i}k_0a }\psi _{-k})
\end{array}
\]
where $m,\, p\in \mathbb{Z}$ are such that $\langle k,\tilde c\rangle =m\pi $ 
and $\langle k,w\rangle =p\pi $. \qquad $\rule{1mm}{1mm}$ 

The relations (\ref{rep2}) define for each $k$ belonging to the set
\begin{equation}
\mathcal{B}_c^{irred}=\{ \ k\in \mathcal{B}_c\backslash \Lambda \ |\ 
\langle k,\tilde c\rangle \not \in \mathbb{Z}\pi \ \  {\rm or}\ \ 
\langle k,w\rangle \not \in \mathbb{Z}\pi  \ \} 
\end{equation}
a two-dimensional irreducible representation $\mathcal{D}_c(k)$.
Some of these representations are equivalent. Particularly, $\mathcal{D}_c(k)=\mathcal{D}_c(-k)$.

\section{Clebsch-Gordan coefficients}
Let $k,\, k'\in \mathcal{B}_c^{irred}$ be such that $k^+=k+k'$ and $k^-=k-k'$ belong to 
$\mathcal{B}_c^{irred}$. The direct product of the representations
$\mathcal{D}_c(k)$ and $\mathcal{D}_c(k')$ admits the decomposition
\[ \mathcal{D}_c(k)\otimes \mathcal{D}_c(k')=\mathcal{D}_c(k^+) \oplus \mathcal{D}_c(k^-) \] 
and the matrices corresponding to $\varrho $, $\sigma $ and $\tau $ are
\[ \fl
\left( \begin{array}{ll}
{\rm e}^{{\rm i}\langle k,\tilde c\rangle } & 0\\
0 & {\rm e}^{-{\rm i}\langle k,\tilde c\rangle }
\end{array} \right)\!\otimes \!
\left( \begin{array}{ll}
{\rm e}^{{\rm i}\langle k',\tilde c\rangle } & 0\\
0 & {\rm e}^{-{\rm i}\langle k',\tilde c\rangle }
\end{array} \right)
\!=\!\left( \begin{array}{cccc}
{\rm e}^{{\rm i}\langle k^+,\tilde c\rangle } & 0 & 0 & 0\\[2mm]
0 & {\rm e}^{{\rm i}\langle k^-,\tilde c\rangle }  & 0 & 0\\[2mm]
0 & 0 & {\rm e}^{-{\rm i}\langle k^-,\tilde c\rangle }  & 0\\[2mm]
0 & 0 & 0 & {\rm e}^{-{\rm i}\langle k^+,\tilde c\rangle } 
\end{array} \right) \]
\[ \fl
\left( \begin{array}{ll}
{\rm e}^{{\rm i}\langle k,w\rangle } & 0\\
0 & {\rm e}^{-{\rm i}\langle k,w\rangle }
\end{array} \right)\!\otimes \!
\left( \begin{array}{ll}
{\rm e}^{{\rm i}\langle k',w\rangle } & 0\\
0 & {\rm e}^{-{\rm i}\langle k',w\rangle }
\end{array} \right)
\!=\!\left( \begin{array}{cccc}
{\rm e}^{{\rm i}\langle k^+,w\rangle } & 0 & 0 & 0\\[2mm]
0 & {\rm e}^{{\rm i}\langle k^-,w\rangle }  & 0 & 0\\[2mm]
0 & 0 & {\rm e}^{-{\rm i}\langle k^-,w\rangle }  & 0\\[2mm]
0 & 0 & 0 & {\rm e}^{-{\rm i}\langle k^+,w\rangle } 
\end{array} \right) \]
\begin{equation}
 \fl 
\left( \begin{array}{ll}
0 & {\rm e}^{{\rm i}k_0a} \\
{\rm e}^{-{\rm i}k_0a } & 0
\end{array} \right)\! \otimes \! 
\left( \begin{array}{ll}
0 & {\rm e}^{{\rm i}{k'}_0a} \\
{\rm e}^{-{\rm i}{k'}_0a } & 0
\end{array} \right)\!=\! 
\left( \begin{array}{cccc}
0 & 0 & 0 & {\rm e}^{{\rm i}k_0^+a } \\[2mm]
0 & 0 & {\rm e}^{{\rm i}k_0^-a }  & 0\\[2mm]
0 & {\rm e}^{-{\rm i}k_0^-a }  & 0 & 0\\[2mm]
{\rm e}^{-{\rm i}k_0^+a }  & 0 & 0 & 0
\end{array} \right)  
\end{equation}
respectively.
The unitary matrix
\begin{equation}
M=\left( \begin{array}{cccc}
1 & 0 & 0 & 0\\[2mm]
0 & 0 & 1 & 0\\[2mm]
0 & 0 & 0 & 1\\[2mm]
0 & 1 & 0 & 0
\end{array} \right) 
\end{equation}
satisfies the relations
{\small 
\[ \fl 
M^{-1}
\left( \begin{array}{cccc}
{\rm e}^{{\rm i}\langle k^+,\tilde c\rangle } & 0 & 0 & 0\\[2mm]
0 & {\rm e}^{{\rm i}\langle k^-,\tilde c\rangle }  & 0 & 0\\[2mm]
0 & 0 & {\rm e}^{-{\rm i}\langle k^-,\tilde c\rangle }  & 0\\[2mm]
0 & 0 & 0 & {\rm e}^{-{\rm i}\langle k^+,\tilde c\rangle } 
\end{array} \right)M= 
\left( \begin{array}{cccc}
{\rm e}^{{\rm i}\langle k^+,\tilde c\rangle } & 0 & 0 & 0\\[2mm]
0 & {\rm e}^{-{\rm i}\langle k^+,\tilde c\rangle }  & 0 & 0\\[2mm]
0 & 0 & {\rm e}^{{\rm i}\langle k^-,\tilde c\rangle }  & 0\\[2mm]
0 & 0 & 0 & {\rm e}^{-{\rm i}\langle k^-,\tilde c\rangle } 
\end{array} \right) \]
\[ \fl 
M^{-1}\!\!
\left( \! \begin{array}{cccc}
{\rm e}^{{\rm i}\langle k^+,w\rangle } & 0 & 0 & 0\\[2mm]
0 & {\rm e}^{{\rm i}\langle k^-,w\rangle }  & 0 & 0\\[2mm]
0 & 0 & {\rm e}^{-{\rm i}\langle k^-,w\rangle }  & 0\\[2mm]
0 & 0 & 0 & {\rm e}^{-{\rm i}\langle k^+,w\rangle } 
\end{array}\! \right)\!\!M\!\!=\!\! 
\left( \! \begin{array}{cccc}
{\rm e}^{{\rm i}\langle k^+,w\rangle } & 0 & 0 & 0\\[2mm]
0 & {\rm e}^{-{\rm i}\langle k^+,w\rangle }  & 0 & 0\\[2mm]
0 & 0 & {\rm e}^{{\rm i}\langle k^-,w\rangle }  & 0\\[2mm]
0 & 0 & 0 & {\rm e}^{-{\rm i}\langle k^-,w\rangle } 
\end{array}\! \right) \]
\begin{equation}
 \fl 
M^{-1}
\left( \begin{array}{cccc}
0 & 0 & 0 & {\rm e}^{{\rm i}k_0^+a } \\[2mm]
0 & 0 & {\rm e}^{{\rm i}k_0^-a }  & 0\\[2mm]
0 & {\rm e}^{-{\rm i}k_0^-a }  & 0 & 0\\[2mm]
{\rm e}^{-{\rm i}k_0^+a }  & 0 & 0 & 0
\end{array} \right) 
M=
\left( \begin{array}{cccc}
0 & {\rm e}^{{\rm i}k_0^+a }  & 0 & 0\\[2mm]
{\rm e}^{-{\rm i}k_0^+a }  & 0 & 0 & 0\\[2mm]
0 & 0 & 0 & {\rm e}^{{\rm i}k_0^-a } \\[2mm]
0 & 0 & {\rm e}^{-{\rm i}k_0^-a }  & 0
\end{array} \right) .
\end{equation} }
\noindent Therefore, the entries of M are Clebsch-Gordon coefficients \cite{K} corresponding to 
the considered direct product. In this case, the only non-null coefficients are
\begin{equation}
(kk'11|k^+1)=(kk'12|k^-1)=(kk'21|k^-2)=(kk'22|k^+2)=1 . 
\end{equation}
More details concerning the Clebsch-Gordon coefficients and their applications in carbon
nanotube physics can be found in the articles of Damnjanovi\'c et al. \cite{D0,D2,Vu}.

\section{Concluding remarks}

The present paper can be regarded as a pure mathematical exercise. We have defined the 
factor sets $\mathcal{L}_c$, the groups $G_c$ acting on $\mathcal{L}_c$ as groups of
permutations, and we have studied certain representations of these groups defined on
some spaces of functions $\psi :\mathcal{L}_c\longrightarrow \mathbb{C}$. We have proved
that the groups $G_c$ are isomorphic to the symmetry groups of single-wall carbon
nanotubes, and the considered representations are directly related to some representations
used in carbon nanotube physics.

The present paper can also be regarded as presenting an alternate mathematical model for 
carbon nanotubes. We think that this alternate approach offers some formal advantages,
namely, certain calculations may be simpler in this approach than in the usual one.

\section*{Acknowledgment}

This  research was supported by a grant CNCSIS.

\end{document}